\documentclass[11pt,twoside]{article}
\usepackage{newpasp}

\usepackage{epsf}
\usepackage{graphicx}

\index{Young, L. M.}

\begin{document}

\title{Molecular Gas in Elliptical Galaxies}
\author{L. M. Young}
\affil{Physics Department, New Mexico Tech, Socorro NM 87801}

% A concise abstract is recommended.  Enter the text of the abstract in
% between the \begin{abstract} and \end{abstract} commands.  Do NOT
% include the word ``Abstract'' in your text; it is insterted
% automatically. Do NOT  make a paragraph break between \begin{abstract} 
% and the first line of the text of the abstract!  Abstracts are required 
% for all papers.

\begin{abstract}
The distribution and kinematics of the molecular gas in elliptical galaxies give 
information on the origin and history of the gas
and the rate of star formation activity in ellipticals.
I describe some preliminary results of a survey which 
will more than double the number of elliptical galaxies
with resolved molecular distributions.
\end{abstract}

% Include keywords if you wish. The keywords.apj file, found on aas.org 
% in the pubs/aastex-misc directory, contains a list of keywords used 
% with the ApJ and Letters.  

%%\keywords{infrared: galaxies -- galaxies: nuclei -- galaxies: starburst}

% That's it for the front matter.  On to the main body of the paper.

\section{Introduction}

Not long ago it became clear that many elliptical galaxies {\it do} have
small amounts of cold interstellar gas and dust (e.g. Knapp et al. 1989;
Goudfrooij et al. 1994). 
The gas and dust have significance far beyond their small masses:
their distribution and kinematics provide
vital clues to the evolution of ellipticals (van Gorkom \& Schiminovich 1997; Sadler et al. 2000).
%For example, disturbed kinematics or non-settled distributions give weight
%to
%the popular idea that the atomic gas was acquired in a merger or
%interaction. (Schiminovich; Sadler)
%The dust lanes have some relation to active Galactic nuclei and jets.
To date, most work on the distribution and kinematics of neutral gas
in ellipticals has focused on the atomic gas.

The present contribution describes a survey to make
maps of the CO emission in ten elliptical galaxies.
The main selection criterion for this sample (after the classification issues, which are admittedly difficult) is the galaxies'
single dish CO line strengths. 
%(e.g. Wiklind, Combes, \& Henkel 1995; Lees et al. 1991; Knapp \& Rupen 1996).
%Galaxies are:  NGC 185, 205, 1275, 807, UGC 1503, NGC 3656, 4476, 5666, 7468,
%4649.
The resolution of the CO data is about 6\arcsec, which is 20 pc (!) for 
the nearest galaxies and 1.5 kpc for the most distant ones.

The goals of the project are to understand (1) the origin of the 
molecular gas in ellipticals
and its prospects for long-term stability;
and (2) how much star formation activity is taking place in ellipticals.
The molecular gas distribution and kinematics
are being compared with the atomic gas, stellar
kinematics, and radio continuum and H$\alpha$ emission.
The sample galaxies are found in 
a variety of environments--- some are isolated, some in loose groups, and
some in the center of Virgo cluster.
They also have a wide range of luminosities; some have active nuclei and 
some show visible signs of recent interactions.

Seven of the ten sample galaxies have already been observed with the BIMA
millimeter interferometer,
and five are clearly detected.
Figure 1 shows the stellar distributions and CO maps for two of them.

\begin{figure}
\includegraphics[keepaspectratio=true,scale=0.3,angle=-90]{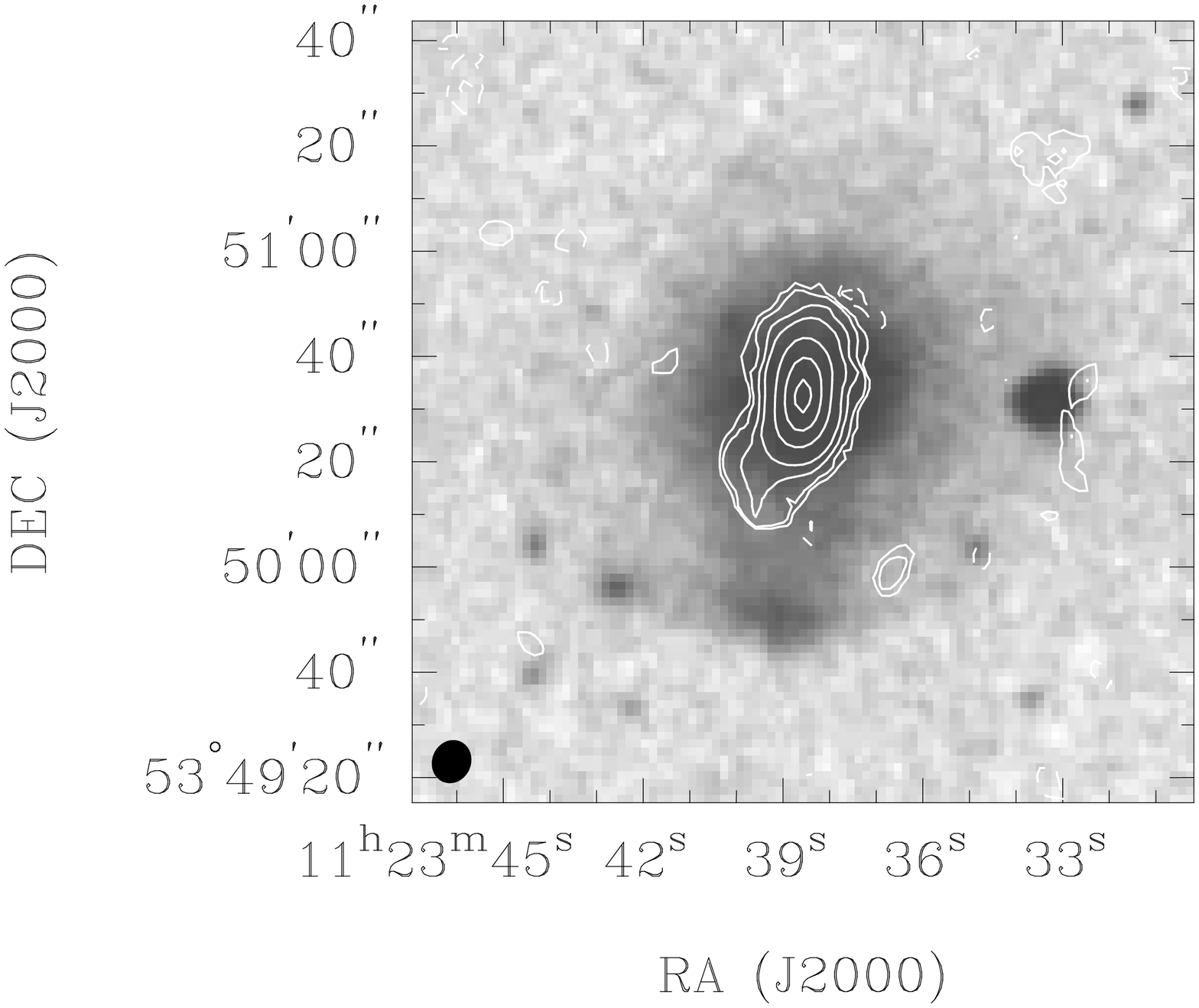}
\includegraphics[keepaspectratio=true,scale=0.3,angle=-90,bb=44 140 580 707]{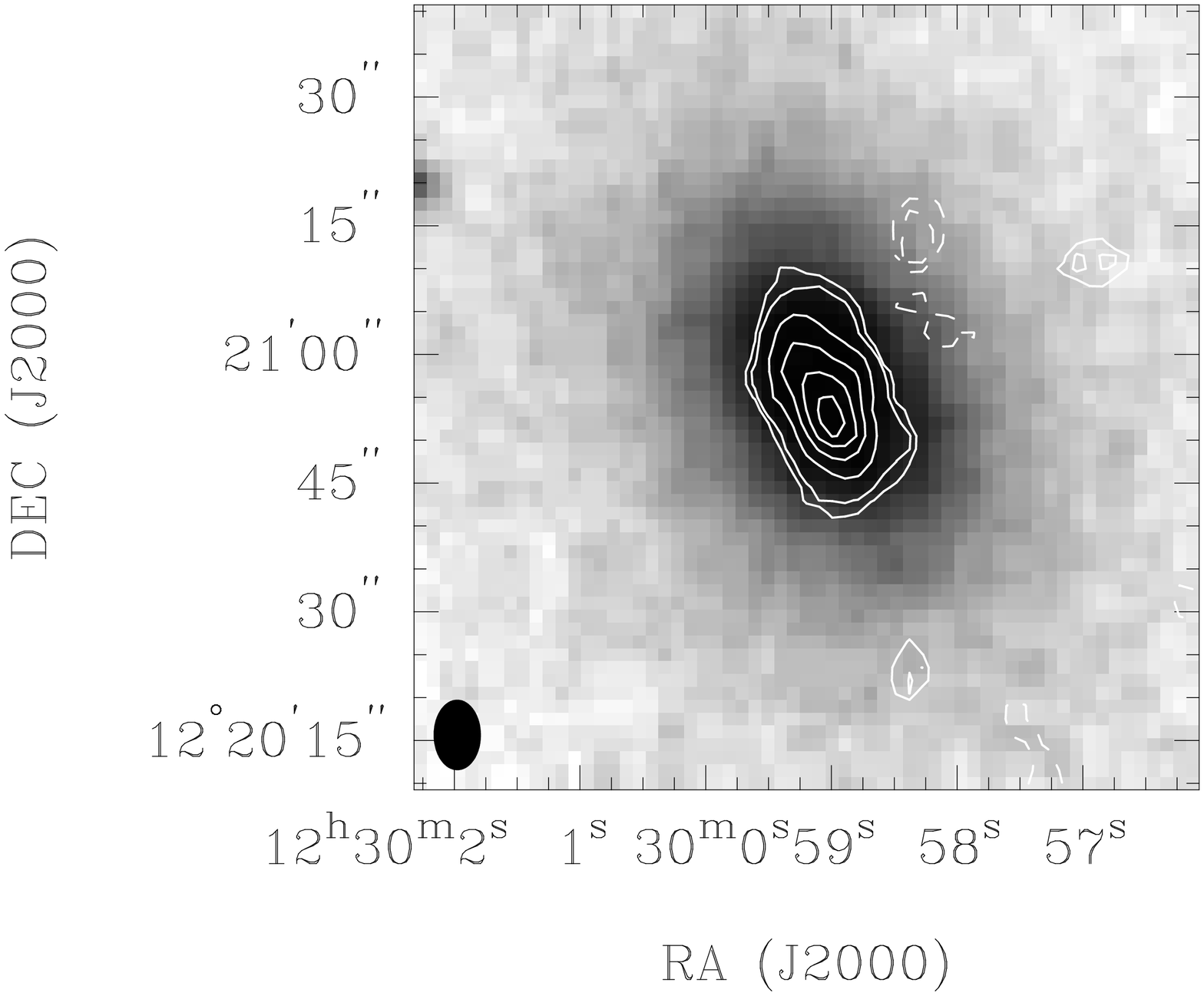}
%\plottwo{3656co+stars.ps}{4476co+stars.ps}
\caption{\small NGC 3656 (left) and NGC 4476 (right).  The greyscales are 
optical images from the Digitized Sky Survey; white contours show the column density
of molecular gas.  NGC~3656 is a merger remnant and NGC~4476 is a companion
of M87 in the Virgo Cluster.
See also Young \& Lo (1996) for NGC~205.}
\end{figure}

% Had some trouble with this plotone and plottwo; fixed it by putting
% the bounding box info at the top of the .ps files instead of the bottom.
% or try this really cool thing from Hibbard's example file:
%  all this fiddling is to get the two parts to have different width
% (plottwo just makes them equal width)
%\begin{figure}[t!]
%\vbox {
%  \begin{minipage}[l]{1.0\textwidth}
%   \hbox{
%       \begin{minipage}[l]{0.39\textwidth}
%       {\centering \leavevmode \epsfxsize=\textwidth 
%        \epsfbox{vla2000_sample_fig2a.ps}}
%       \end{minipage} \  \hfill \
%       \begin{minipage}[r]{0.61\textwidth}
%       {\centering \leavevmode \epsfxsize=\textwidth 
%        \epsfbox{vla2000_sample_fig2b.ps}}
%       \end{minipage} \  \hfill \
%  }
%  \end{minipage} \  \hfill \
%  \begin{minipage}[l]{1.0\textwidth}
%\caption{\small Examples from the HI Rogues Gallery.
%Left: VLA C+D array HI observations of NGC 4038/9. From Hibbard, van der
%Hulst \& Barnes 2000. Right: VLA C-array observations of NGC 5506/7, 
%from Mundell 1997.}
%  \end{minipage}
%}
%\label{fig2}
%\vspace*{-0.4cm}
%\end{figure}

%  See also gr8hifigs.tex for use of \includegraphics and, I think,
%the POWRE proposal.

\section{Some Initial Results}

Except for NGC~185 and NGC~205, the smallest members of the sample,
the molecular gas is distributed in a smooth, regular
disk in solid body rotation.
There is a strong asymmetric warp in the CO of NGC~3656, but otherwise the disks
are not disturbed.
In two cases the galaxies' stars are known to have
large velocity dispersions and small rotational velocities;
the CO rotation is {\it consistent} with that of the stars.
%One interesting case of a stellar counterrotating core, which is
%also counterrotating with respect to the CO.
Usually the HI kinematics are found to be inconsistent with those
of the stars, which is taken as evidence for an external gas origin. 
Perhaps the situation is more complicated for the molecular gas.

A remarkable result comes from comparing the molecular gas
distribution in NGC~5666 to the radio continuum emission
in that galaxy (Wrobel \& Heeschen 1988).  
The radio continuum emission is extended; it has exactly the same extent
as the resolved CO emission. 
Thus,
%Since molecular gas is the raw material for star formation,
the radio continuum and FIR emission can be attributed
to star formation (3 M$_\odot$ per year) rather than to an active galactic nucleus or to an
old stellar population.
% should give a reference here.
The other galaxies in the sample will help to show whether
this situation is common or uncommon.

Comparisons of the molecular gas properties and the
star formation activity in the different galaxies (isolated vs. in clusters,
giant vs. dwarf, and so on)
will allow us to probe elliptical galaxy evolution as a function of
galaxy properties and/or environment.

% For examples on including figures, see the file vla2000_sample.ps
% at http://www.nrao.edu/vla2000/proceedings/. 
% For examples of figures, equations or tables, please see the file
% vla2000_man.ps at the same site. Also available as
% newpaspman.ps at http://www.aspsky.org/pubs/authors.html

% comment this out if you want to include acknowledgements

%\acknowledgements

\end{document}